\documentstyle[11pt,newpasp,twoside,psfig]{article}
\def\edcomment#1{\iffalse\marginpar{\raggedright\sl#1\/}\else\relax\fi}
\reversemarginpar

\begin{document}

\title{The Dispersal of Young Stars and the Greater Sco-Cen Association}

\author{Eric E. Mamajek$^1$ and Eric D. Feigelson$^2$}

\affil{$^1$ Department of Astronomy, Steward Observatory, University
of Arizona, 933 N Cherry Ave, Tucson AZ 85721 USA}

\affil{$^2$ Department of Astronomy \& Astrophysics, Pennsylvania State 
University, University Park PA 16802 USA}

\begin{abstract}

We review topics related to the dispersal of young stars from their
birth-sites, and focus in particular on the entourage of young stars 
related to the ongoing star-formation event in 
the Sco-Cen OB association. We conduct a follow-up kinematic study
to that presented in Mamajek, Lawson, \& Feigelson (2000; ApJ 544, 356) 
amongst nearby, isolated, young stars. In addition to the $\eta$
Cha and TW Hya groups, we find several more intriguing Sco-Cen
outlier candidates: most notably $\beta$ Pic, PZ Tel, HD 199143, and HD 100546. 
We discuss the connection between Sco-Cen and the southern 
``150 pc Conspiracy'' molecular clouds, and in
particular, Corona Australis. The kinematic evidence suggests that many
of the nearby, isolated $\sim$10 Myr-old stars 
were born near Sco-Cen during the UCL and LCC starbursts 10--15 Myr ago.
We hypothesize that these stars inherited 5--10 km s$^{-1}$ velocities moving away
from Sco-Cen, either through molecular cloud turbulence, or through
formation in molecular clouds associated with the 
expanding Sco-Cen superbubbles (e.g. Loop~I).

\end{abstract}
A question arose several times at the YStars conference:  
``Why are nearly 
all of the nearby, very young stars located in the southern, rather 
than northern, sky?''
The ``Gould Belt'' is a general answer, however a more specific
answer may involve the nearest, large component of the belt: Sco-Cen.
We suggest that the Sco-Cen Association (Sco OB2) 
is largely responsible for this asymmetry through two mechanisms: 
the dispersal of stars that formed with the main Sco-Cen OB subgroups in the 
same giant molecular cloud (GMC) complex; and the concentration of gas and 
subsequent star-formation initiated by the Sco-Cen superbubbles. 

We present our heuristic argument in several steps.  First, we provide 
an historical perspective (\S 1) and explore mechanisms of the dispersal of 
pre-main sequence (pre-MS) stars from their natal clouds (\S 2).  In \S3, we 
discuss observations (kinematics, ISM, etc.) linking young stars and groups
to the Sco-Cen OB association.   Finally, we discuss the possible relationship  
between the Sco-Cen superbubble and regions of current star formation (\S 4).  
Many of these thoughts are built on foundations laid in our earlier studies 
(Feigelson 1996; Mamajek, Lawson, \& Feigelson 2000, MLF00).

\section{The missing population of older pre-main sequence stars}
Stars are formed in molecular clouds which contain most of the mass but 
occupy only a tiny portion of the volume of the Galactic disk.  Their 
dispersal into the general star field was historically attributed to 
relatively slow 
processes: the evaporation of open clusters, and gravitational scattering 
processes that produce the observed correlation between stellar age and 
Galactic scale height.  Early studies of the kinematics of pre-MS stars reported 
radial velocities within 1\,km\,s$^{-1}$ of their natal clouds, with a few exceptions 
like FK Ser (Herbig 1973), so there was little evidence for early dispersal. 

But there was a puzzle: for many years, stars in the intermediate stage of 
stellar evolution between the classical T Tauri stars and the Zero-Age Main 
Sequence could not be found near molecular clouds (Herbig 1978).  They 
obviously must exist in considerable numbers, just as there must be an order 
of magnitude more children of ages $1-10$ yr old than there are infants with 
ages $<1$ yr in any steady-state human population.  

A significant portion of this missing population of  intermediate-aged pre-MS 
stars was finally found among the $\sim$10$^5$ sources of the {\it ROSAT} 
All-Sky Survey (RASS).   Unlike infrared excesses or prominent optical 
emission lines, X-ray emission is elevated $10^1-10^4$ above main 
sequence levels for the entire duration of the pre-MS phases, and thus provides 
an excellent criterion to distinguish older pre-MS stars from the general Galactic 
star population (Feigelson \& Montmerle 1999).   In an important series of 
papers, {\it ROSAT} scientists spectroscopically identified dozens of lithium-
rich, magnetically active weak-lined T Tauri stars (Neuh\"auser 1997, and
references therein).  
Unlike traditional T Tauri samples that were spatially 
concentrated around molecular clouds, the young RASS stars typically lie 
several to tens of parsecs from any active star forming region. For our
discussions, recall that 1 km\,s$^{-1}$ translates into 1.0\,pc\,Myr$^{-1}$. 

Nearly 10$^4$ RASS sources are associated with stars in
the Tycho astrometric catalog ($m_V$ $<$ 11), and many are concentrated along the 
Gould Belt of molecular clouds and OB associations (Guillout et al. 
1998).  When X-ray flux limits and the full extent of the 
Gould Belt are taken into account, the RASS sources represent a population of 
tens of thousands of young stars.  Clearly many of the missing older pre-MS stars 
have now been located.

\section{Theoretical origins of dispersed young stars}
This discovery of so many widely dispersed X-ray-emitting pre-MS stars 
requires that much of the stellar dispersal from active star-forming regions 
occurs on timescales of $10^6-10^7$ yr. Several ideas have emerged to
explain this phenomenon.

\begin{enumerate}
\item Stars may undergo gravitational scattering during their protostellar or 
T Tauri phases resulting in the ejection of some stars at high velocities 
(Sterzik \& Durisen 1995; Reipurth 2000; Sterzik, this volume).  This 
process preferentially occurs in triple systems where a hard binary ejects the 
third member.  This process has specific predictions.  The run-away pre-MS 
stars should: a) be single and not in multiple systems; b) have a mass 
function biased towards low masses, with many brown dwarfs; and 
c) have space motions that point radially from known modern star-forming regions.  
There is little evidence that the dispersed RASS young stars have these 
properties.   As the dynamical arguments are convincing, it 
seems likely that dynamically ejected stars will be found with the 
deeper astrometric surveys such as {\it FAME} (Greene, this volume) and 
{\it GAIA}.  
\item  Rather than ejecting stars from molecular clouds, pre-MS stars may 
appear isolated because their molecular clouds disappear shortly after 
star formation occurs (e.g. Palla \& Galli 1997).  There is evidence that star 
formation occurs over a short timescale within a given molecular cloud 
(Elmegreen 2001) and, if the cloud is quickly dispersed by stellar 
radiation and outflows, the resulting pre-MS stars would appear dispersed among 
older Galactic populations.  This model predicts frequent formation and 
dissolution of molecular clouds and near-coevality of the stellar 
populations produced ($<${\it few} Myr). 

\item  Pre-MS stars may be widely dispersed because they inherit the motions 
of their natal molecular gas, which exhibits supersonic motions on large 
scales in GMCs (Feigelson 1996). Groups of stars from
different parts of a GMC will inherit motions from cloud turbulence and
will be dispersed over tens of parsecs during their pre-MS evolution. 
This model is supported 
by a wealth of observational evidence for velocity dispersions of order $5-
10$ km\,s$^{-1}$ on scales of $20-100$\,pc in molecular cloud complexes.  For 
example, CO maps of the Ophiuchus clouds, with $10^4$ M$_{gas,\odot}$  
and $10^2$ M$_{stars, \odot}$ of young stars, and Orion clouds, with 
$10^5$ M$_{gas,\odot}$ and $10^3$ M$_{stars,\odot}$ show radial 
velocity gradients of $6-11$ km s$^{-1}$ (e.g. de Geus et al.\ 1990; Tatematsu et 
al.\ 1993; Miesch et al.\ 1999).  In the larger clouds, these high-velocity 
motions are shared by the massive cores where rich star clusters form, 
as well as smaller cores and cloudlets.  Truly giant cloud complexes, like 
W51 with $10^6$ M$_{gas,\odot}$  and $10^4$ M$_{stars,\odot}$,  show velocity 
dispersions around 20 km s$^{-1}$ on scales of 100\,pc 
(Carpenter \& Sanders 1998).  Such high velocities are ubiquitous in 
molecular cloud complexes, and are summarized by the relationship $\Delta 
v$ $\simeq$ 5 km\,s$^{-1}$\,(r/50\,pc)$^{1/2}$ (Larson 1981; Efremov \& Elmegreen 
1998).  This relation and other fractal properties of molecular clouds are 
commonly interpreted as manifestations of supersonic MHD turbulence 
(V\'azquez-Semadeni et al.\ 2000). 
\end{enumerate}

Given the evidence for supersonic gas motions, it is difficult to 
imagine that many pre-MS stars born in a GMC complex would 
{\it not} be endowed with 5--10 km\,s$^{-1}$ velocities upon birth 
(Feigelson 1996).  
Members of a cluster produced by a massive molecular core with a small 
velocity dispersion will appear as an easily-found rich star association.  
But it will be surrounded by an inhomogeneous halo of sparser star clusters 
produced in smaller molecular cores.  After 10 Myr, the dispersed star 
complex may be spread over $\sim$100\,pc with 
space motions convergent to a 20--50\,pc region representing the parental 
molecular cloud.  

\section{Are the dispersed young, nearby stars Sco-Cen outliers?}
The Sco-Cen OB Association provides a unique laboratory for evaluating the
dispersal of young stars. Sco-Cen (Sco OB2) is the nearest region of 
recent and on-going high-mass star-formation to the Sun (see review
by Blauuw 1991), and its projected boundaries encompass $\sim$5\% of the sky 
(de Zeeuw et al 1999; dZ99).
Sco-Cen consists of 3 OB subgroups: 
Upper Sco (US; 5--6 Myr; 145\,pc), Upper Centaurus Lupus (UCL; 14--15 Myr;
142\,pc),
Lower Centaurus Crux (LCC; 11--12 Myr; 118\,pc; de Geus et al. 1989; dG89). 
de Geus (1992; dG92) summarized the studies of the early-type stellar populations of
Sco-Cen as well as their effects 
on the local interstellar medium (winds \& supernovae; later discussed in \S 4). 

In our recent study (MLF00), we present an argument 
connecting the origins of three dispersed pre-MS stellar groups to 
Sco-Cen: the compact $\eta$ Cha cluster ($d$ = 97\,pc, age $\simeq$ 8 Myr; 
Mamajek, Lawson, \& Feigelson, 1999, Lawson, this volume); 
the extended TW Hydra Association (TWA, $d$ $\simeq$ 
55\,pc, age $\simeq$ 10 Myr, see Webb, this volume); and a small group 
of stars near $\epsilon$ Cha ($d$ $\simeq$ 110\,pc, age $\approx$ 5--15 Myr; MLF00). 
Their space motions are moving nearly radially away from the
two oldest Sco-Cen OB subgroups and their ages are consistent with the 
times of minimum separation from the subgroups.

\subsection{A search for new candidate outliers of Sco-Cen}

We suggest that $\eta$ Cha, $\epsilon$ Cha and the TWA are only part of a
dispersed population of Sco-Cen outliers formed in/near the Sco-Cen GMC 
5--15 Myr ago.  Some of these outliers should lie, by chance, quite near the Sun 
and thus constitute excellent targets for the study of pre-MS stellar and disk 
evolution.   To pursue this idea, we compiled from the literature a 
heterogenous list
of other young dispersed Li-rich dwarfs, Herbig Ae/Be stars,
and disked early-type stars in the solar neighborhood.  Our primary
astrometric data sources are the {\it Hipparcos} (ESA 1997) and
Tycho-2 catalogs (Hog et al.\ 2000), and we take radial velocities
($v_R$) from the compilations of Barbier-Brossat et al. (2000) or 
Maloroda et al. (2001), unless otherwise noted.
We calculate and compare 
the motions of these stars to Sco-Cen as described in MLF00, assuming
constant linear motion over the past $\sim$15 Myr (i.e. over the
known star-formation history of Sco-Cen). We considered an outlier
candidate to be a star which was within $\sim$50\, pc of a Sco-Cen 
subgroup within the past 15 Myr. We omit the Galactic
potential, since this interval is much smaller than the epicyclic 
($\kappa$) and vertical ($\nu$) periods. We claim this is satisfactory
for this initial effort. The dangers in extrapolating stellar
motions back in time are nicely summarized by Soderblom et al. (1990), 
however most of the objections pertain to longer intervals and older
groups ($>$100 Myr). Here, we report a few interesting cases:

$\bullet$ {\bf HD 100546} (B9Ve; $\ell$, $b$ = 296.4$^{o}$, --8.3$^{o}$; 
$d$ $\simeq$ 103\,pc) is an isolated Be star apparently associated with 
the dark cloud DC 296.2--7.9 with a lower-limit age of $>$10 Myr 
(van den Ancker et al.\ 1998). dZ99 list it as an LCC member by virtue 
of its Hipparcos astrometry, and it lies on the southern periphery of the 
subgroup. HD 100546 is very close (0.6$^{\circ}$) to a co-distant, co-moving star: 
{\bf HD 101088} (F5IV; $d$ $\simeq$ 101\,pc).
Vieira et al.\ (1999) claim HD 101088 is an unrelated foreground star, 
however dZ99 list it as another LCC member. 
We calculate a space motion of U,V,W = (--10.5, --19.7, 
--7.7) $\pm$ (1.1, 1.0, 0.5)\,km\,s$^{-1}$ for HD 100546, which is only 
2\,km\,s$^{-1}$ from the motion of LCC. HD 101088 should have an 
identical $v_R$ as HD 100546 to confirm its status as a pre-MS LCC 
member. 

$\bullet$ {\bf 51 Oph} (A0Ve; $\ell$, $b$ = 2.5$^{\circ}$, +5.3$^{\circ}$; 
$d$ $\simeq$ 131 
pc) is one of the most unusual disked Herbig Ae stars yet identified, and
its status as pre-MS vs. post-MS is still unclear (van den Ancker et 
al.\ 2000a). With a U,V,W space motion of (--9.8, --12.3, --12.3) $\pm$ 
(2.4, 1.3, 1.4)\,km\,s$^{-1}$, 51 Oph was within $\sim$20\,pc of US some 4 
Myr ago. This is similar to the age of the US subgroup (dG89). It is moving 
away from US at 9 km\,s$^{-1}$, and we consider it an US outlier candidate.

$\bullet$ {\bf HD 199143} (F8V; $\ell$, $b$ = 30.4$^{\circ}$, --35.0$^{\circ}$; $d$ 
$\simeq$ 48\,pc) van den Ancker et al.\ (2000b, this volume) describes 
this F8V star as an isolated $\sim$10 Myr-old star with a T Tauri-type 
companion, {\bf HD 358623}. They report 
$v_R = -9 \pm 16$ km\,s$^{-1}$ for HD\,199143, giving it a space motion of 
(--11.0, --15.8, --9.2)\,$\pm$\, 
(11.3, 6.7, 9.2)\,km\,s$^{-1}$.
We investigate the motions of HD 199143 for $v_R$ values within the error bars, 
and find the star was within 40\,pc of the UCL subgroup $\sim$13 Myr ago for 
$v_{R}\,\simeq\,-9\,\pm\,2$\,km\,s$^{-1}$. With that $v_R$, the HD 199143 system 
is an UCL outlier candidate, moving away from UCL at $\sim$10\,km\,s$^{-1}$. 

$\bullet$ {\bf V824 Ara} (K1Vp; $\ell$, $b$ = 324.9$^{\circ}$, --16.3$^{\circ}$; $d$
 $\simeq$ 31\,pc) Pasquini et al. (1991) made a strong case for this star being 
pre-MS instead 
of post-MS, and Strassmeier \& Rice (2000) quote an age 
of 18 Myr. With a space motion of (--7.9, --18.9, --10.3) $\pm$ 
(1.0, 0.9, 0.5)\,km\,s$^{-1}$, we find that V824 Ara was within $\sim$25\,pc of UCL
$\sim$12 Myr ago, and is moving away from UCL at 7 km\,s$^{-1}$.

$\bullet$ {\bf V343 Nor} (K0V; $\ell$, $b$ = 323.8$^{\circ}$,  --1.8$^{\circ}$; $d$ $\simeq$ 
40\,pc). This is a nearby young, magnetically active star discussed by Anders et 
al.\ (1991) as a ``Local Association'' member. Its Li $\lambda$6707 EW of 302 
m\AA\/ suggests an age younger than IC 2602 ($<$30--50 Myr). We calculate
a space motion of ($-10.8, -16.8, -10.2$) $\pm$ 
(1.0, 0.9, 0.5)\,km\,s$^{-1}$, which places V343 Nor about $\sim$40\,pc from the 
UCL subgroup $\sim$9 Myr ago. V343 Nor is moving away from UCL at about 10 km\,s$^{-1}$.

$\bullet$ {\bf PZ Tel} (K0Vp; $\ell$, $b$ = 346.2$^{\circ}$,  --20.8$^{\circ}$; $d$ $\simeq$ 
50\,pc). Favata et al.\ (1998) estimate the age of this well-studied 
active dwarf to be $\sim$20 Myr old. Using the space motion from Barnes et al.\ 
(2000), we find that PZ Tel was within $\sim$40\,pc of UCL $\sim$13 Myr ago.  PZ 
Tel is moving away from UCL at 8 km\,s$^{-1}$. 

$\bullet$ {\bf $\beta$ Pic} (A5V; $\ell$, $b$ = 258.4$^{\circ}$, 
--30.6$^{\circ}$; $d$ $\simeq$ 19\,pc). Barrado \& Navascues et al. 
(1999; B\&N99) 
estimated an age of 20\,$\pm$\,10 Myr for this prototype disked A star,
and we use their U,V,W vector in our analysis. The motion of $\beta$ Pic 
is $\sim$7 km\,s$^{-1}$ directed away from UCL, and it lay 
within $\sim$50\,pc of UCL about $\sim$13 Myr ago.  B\&N99 find that
the two active M dwarfs {\bf GJ 799} and {\bf GJ 803} are comoving
with $\beta$ Pic, and that they are 2 of the 3 most active 
(youngest) nearby M dwarfs out of a sample of 160. 
Using the B\&N99 space motions, we find that GJ 799 was $\sim$70\,pc 
from UCL $\sim$11 Myr ago, and GJ 803 was $\sim$60\,pc from 
UCL $\sim$12 Myr ago. The kinematic results and age agreement are 
very suggestive of a 
collective Sco-Cen origin, but not overwhelmingly convincing. Much of the 
separation between these stars and UCL $\sim$12 Myr ago was in the Z 
direction, so a future investigation including the Galactic potential will
better address the plausibility of the $\beta$ Pic group 
(and many of the other candidates mentioned) as being related to the 
Sco-Cen star-formation event. If the trio are true Sco-Cen outliers, they
should have ages of $<$13 Myr; which is at the low end of recent age
estimates (Barrado \& Navascues, this volume). A lithium depletion study
of the M dwarfs could further address the issue of their ages.

A few other stars appear to be moving away from 
Sco-Cen very quickly ($\sim$15 km\,s$^{-1}$ for LQ Hya, $\sim$25 km\,s$^{-1}$
for GJ 182 and EQ Vir) but their lithium abundances suggest they are older
than the Sco-Cen subgroups; hence they are clearly unrelated.  
Other young, active stars (e.g. AB Dor, HD 105, V383 Cen, etc.) have motions and 
ages completely inconsistent with a Sco-Cen origin. Many could be older Gould Belt 
stars ($\sim$50 Myr), or even young MS disk stars. 
Here we mention some stars that appear to {\it not} be kinematically 
tied to Sco-Cen:

$\bullet$ {\bf FK Ser} (K5Vp; $\ell$, $b$ = 20.3$^{\circ}$, 
+2.2$^{\circ}$; d $\sim$ 100\,pc) is a binary isolated T Tauri star
discussed by Herbig (1978). The star has a poor {\it Hipparcos} parallax
($\pi$ = 9.42\,$\pm$\,6.17 mas), and we adopt the long-baseline
Tycho-2 proper motion which is 2--3$\times$ better than Hipparcos'. 
From Herbig (1973) and Zappala (1974), 
we calculate a mean $v_R$ = --10\,$\pm$\,3 km\,s$^{-1}$. 
Calculating space motions within the 1-$\sigma$ errors for the 
$\pi$ and $v_R$ in steps of 1 mas and 1 km\,s$^{-1}$, 
we find no pass near Sco-Cen in the past - hence we do not consider
an outlier candidate.

$\bullet$ {\bf HD 163296} (A2Ve; $\ell$, $b$ = 7.2$^{\circ}$, +1.5$^{\circ}$;
 $d$ $\simeq$ 122\,pc) is an isolated Herbig Ae star recently found to have both a 
disk and outflow (Grady et al.\ 2000).
van den Acker et al.\ (1998) found an age of 4$^{+6}_{-2.5}$ Myr
from isochrone fitting. Spatially the star is within 10$^{\circ}$ of, 
and nearly co-distant with the $\rho$ Oph streamers oriented away from US.
We calculate a space motion of 
(--0.9, --22.9, --7.8) km\,s$^{-1}$,
which has the star moving {\it towards} the Sco-Cen subgroups.
Despite its proximity to Ophiuchus, we can easily rule out common proper
motion with Upper Sco (its nearest Sco-Cen subgroup), and radial motion
away from any of the subgroup.

$\bullet$ {\bf HD 17925} (K1V; $\ell$, $b$ = 192.1$^{\circ}$,  --58.3$^{\circ}$; 
$d$ $\simeq$ 
10\,pc). This was the original ``Sco-Cen outlier'' proposed by Cayrel de Strobel 
\& Cayrel (1989).  HD 17925 came no closer than $\sim$80\,pc of Sco-Cen in
the past. For its spectral type, and Li $\lambda$6707\AA~equivalent width  
of 200 m\AA~(Favata et al.\ 1997), an age of $\sim$100 Myr would be more 
appropriate -- hence it is unrelated to Sco-Cen.

To summarize our study, we find that several of the extremely
young, nearby stars appear to be moving away from Sco-Cen, and have ages
younger than or similar to that of the Sco-Cen OB subgroups. Many are not
surprisingly, unrelated to this star-formation complex. In {\bf Fig. 1} 
we plot the motions of the Sco-Cen outlier candidates over the past
12 Myr. The motions are 
with respect to the oldest Sco-Cen subgroup (UCL; heliocentric
U,V,W = --3.9, --20.3, --3.4) km\,s$^{-1}$. Note that the ages of some
of the groups are younger than 12 Myr (i.e. US, CrA), and but most of
the times of closest pass are $\sim$12-13 Myr ago.

\subsection{The TW Hya Association and Lower Centaurus Crux}

The TWA members with {\it Hipparcos} astrometry 
in Webb et al.\ (1999) had U,V,W vectors
consistent with an origin near the LCC OB subgroup some $\sim$10 Myr ago (MLF00). 
The ages of TWA 
($\sim$10 Myr) and LCC (11-12 Myr) are nearly identical.  
{\it All} of the new TWA members reported by Zuckerman et al.\ (2001) 
lie within the projected 
boundaries of the LCC subgroup (dZ99), and have estimated distances 
(d $\simeq$ 70--100\,pc) even closer to LCC than the original TWA stars 
({\bf Fig. 2}). The proper motions ($\mu$) are shown in Fig. 2, but it is mostly
solar reflex motion. 
Fig. 2 also illustrates that searching for new TWA
members will uncover increasing numbers of
T Tauri stars at southern declinations ($\delta$ $<$ --40$^{\circ}$), 
but these will most likely be bona fide LCC members.
A specific example is TWA 19 (HIP 57524) with a {\it Hipparcos} distance of 
104$^{+18}_{-13}$\,pc,  which dZ99 lists as a kinematic member of LCC.
At a mean distance of 118\,pc, the projected width of the LCC subgroup 
($\sim$25$^{\circ}$) corresponds to a depth of $\sim$50\,pc;  indeed many of the 
early-type members are closer than 100\,pc (dZ99).  Extrapolating from
the number
of B stars with a typical initial mass function (IMF), dG92 suggests that the 
total stellar population of LCC is $\sim$1300; most of which 
will be T Tauri stars. A spectroscopic survey 
of late-type RASS-Tycho stars has uncovered dozens of LCC T Tauri 
stars with kinematic distances $<$100\,pc (Mamajek et al., in prep.). 
The continuity in the distances of TWA and LCC stars, and their coevality
lead us to conclude that there is no clear division between 
TWA ($\sim$10$^{1.5}$ stars) and LCC ($\sim$10$^{3}$ stars). 
Despite their nearly coherent motions, it is unclear
whether the TWA stars formed individually in isolation (under similar circumstances) 
or originally from a small cluster analagous to $\eta$ Cha which has 
evaporated. Makarov \& 
Fabricius (2001) report an expansion age of 8 Myr for the TWA; however
the significance of this measurement is clouded by the inclusion of 
potential interlopers
(i.e. 23/31 of their TWA kinematic members have yet to be confirmed as
pre-MS, but were included in the analysis). They, too, conclude that the division 
between LCC and TWA is tenuous at best.

\begin{figure}
\centerline{\psfig{figure=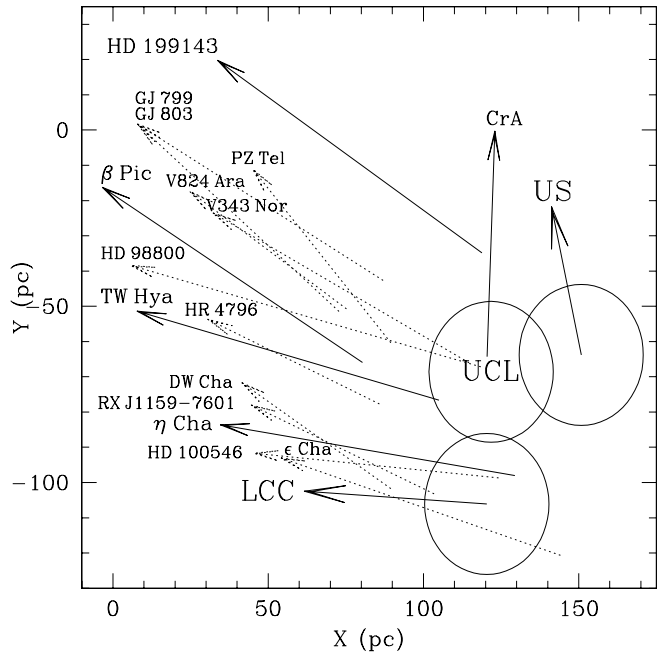,width=8.6cm,height=8.6cm,angle=0}}
\caption[]{The motions of nearby young stars 
over the past 12 Myr, with respect to the 
UCL OB subgroup of Sco-Cen (here, stationary). 
The current position is at the arrow tip, and the unlabelled end is 12 Myr ago.
The 40\,pc circles encapsulate the OB subgroups 12 Myr ago.}
\end{figure}

\begin{figure}
\centerline{\psfig{figure=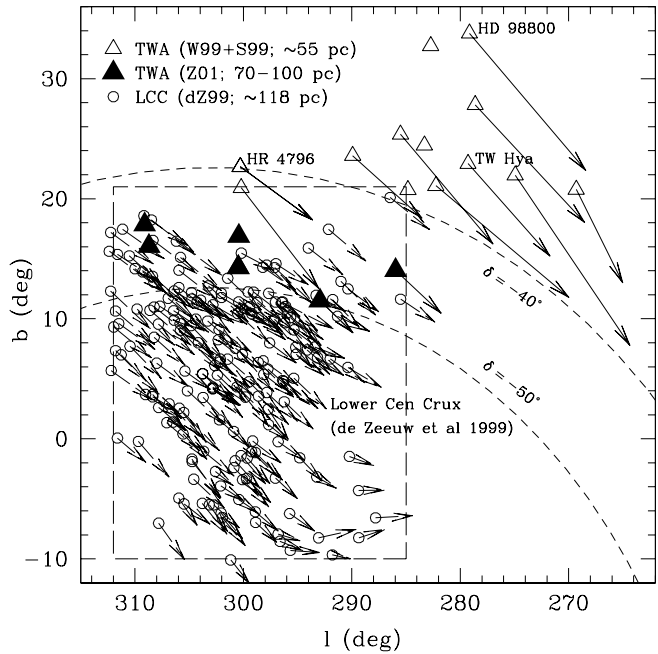,width=8.6cm,height=8.6cm,angle=0}}
\caption[]{The TW Hya association and the Lower Cen Crux OB subgroup.
TWA 1-13 are from Webb et al. (1999) and Sterzik et al. (1999), while
TWA 14-19 are from Zuckerman et al. (2001) (see \S 3.2)}
\end{figure}

\section{Sco-Cen and the southern ``150\,pc Conspiracy'' 
molecular clouds}

Aside from Taurus clouds, it is well known that most of the nearest molecular cloud 
complexes lie in the southern hemisphere at a distance around 150\,pc.  
Simple application of the initial mass function and stellar evolution indicates 
that the Sco-Cen subgroups must have had several ($\sim$10) O and early-B 
stars that have evolved and gone supernovae.  The stellar winds and 
supernova remnants of Sco-Cen O and early-B stars have had a major impact 
on the surrounding interstellar medium and the progenitor GMC. Using 
superbubble theory (McCray \& Kafatos 1987, Tenorio-Tagle \&
Bodenheimer 1988), dG92 argue that the energy input from the massive 
stars in the OB subgroups produced the observed network of superbubbles 
some $\sim$100\,pc in radius centered on Sco-Cen. The largest of
these is Loop I, which is centered on the UCL subgroup.

These superbubbles are seen as a series of arcs of H I and warm dust towards
the 4th Galactic quadrant, and encompassing
$\sim$20\% of the sky. 
The high galactic latitude molecular clouds in the southern hemisphere 
appear to be preferentially placed along the edge of Loop I (Gir et al. 1994).
The total H I mass of the superbubbles is $\sim$5\,$\times$\,10$^5$ M$_{\odot}$, 
which is likely composed of gas swept up from 
the ambient ISM and the Sco-Cen progenitor GMC. The bubbles have 
expansion velocities of $\sim$10 km\,s$^{-1}$ and a total kinetic energy 
of $\sim$10$^{51}$\,erg (dG92). 

The superbubbles may be responsible for triggering star formation in some or 
most of the nearby star-forming regions in the southern sky such as Lupus, 
Ophiuchus, Corona Australis, and Chamaeleon. The Musca and Coalsack clouds
may be in the earliest stages of star-formation. 
All of these lie 100$-$200\,pc from the Sun in the vicinity of the Sco-Cen subgroups.
A review of observations and theory relating to the triggering of star-formation in
swept-up neutral shells is reviewed by Elmegreen (1998). It has been argued 
that the ionization front and 
winds of the massive stars in US is now producing a shock front that is compressing 
molecular material in the core of the Ophiuchi cloud, and is blowing long 
cometary tails in the outer part of the cloud (e.g. Loren et al. 1989). 
The Lupus clouds lies on the 
edge of the young superbubble around US (dG92), while the CrA molecular cloud is 
embedded within the thin H\,I shell associated with the Loop I superbubble (\S 5).  
The Coalsack, Musca, and Chamaeleon clouds have also been physically
associated with each other through an extended interstellar dust ``sheet'',
which appears to be the thin shell of the Loop I superbubble 
(Corradi et al. 1997). 
From the ISM studies mentioned previously, 
it appears likely that the southern ``150\,pc''
clouds are linked to the on-going Sco-Cen star-formation episode. 

\section{The Corona Australis Pre-MS cluster: a missing link?}

The Corona Australis (CrA) molecular cloud is an active star-forming cloud 
complex with a compact cluster of protostars (the ``Coronet'') surrounded by 
an extended group of T Tauri stars (Neuh\"auser et al.\ 2001, and refs. therein). 
We calculate the space motion of the CrA group to be (U,V,W) = (--3.7, 
--15.0, --6.9) km\,s$^{-1}$ using the mean radial velocity from 
Walter et al.\ (1994) ($-1.4\,\pm\,0.6$ km\,s$^{-1}$), 
the weighted mean $\mu$ from several Tycho-2 entries (S CrA, 
TY CrA, HBC 676, HBC 678, and CrAPMS 4SE) ($\mu_\alpha$, 
$\mu_\delta$) = (+2.9\,$\pm$\,0.8, --27.4\,$\pm$\,0.9) mas yr$^{-1}$, and the 
distance obtained by Casey et al.\ (1998) for the eclipsing binary TY Cra 
(129\,$\pm$\,11\,pc). The stars in the CrA cloud are mostly $<$3 Myr old. 
Tracing the 
motion of the CrA group back, we find that it lay within $\sim$30\,pc of UCL 
some 14 Myr ago. The CrA complex is moving radially away from UCL at 
7 km\,s$^{-1}$, similar to the expansion velocity of the Sco-Cen
superbubbles ($\sim$10 km\,s$^{-1}$; dG92). Cappa de Nicolau \& P\"oppel (1991) 
and Harju et al. (1993) found corroborating evidence that the CrA clouds 
are embedded within the massive H\,I shell of Loop I. 
We conclude that the CrA stars have inherited the motion of the
CrA molecular cloud, which itself is embedded in the Loop I shell, and
the whole complex is moving radially away from the $\sim$14--15 Myr-old 
UCL OB subgroup. The superbubble provides a natural mass and 
momentum source for forming the CrA molecular clouds and its associated
stars.

\section{Summary}

The question  ``Why are nearly all of the very young, nearby stars located in 
the southern, rather than northern, sky?''  may have a simple answer: "Because 
of the Sco-Cen Association".   The Sco-Cen Association likely has a total 
population of $\sim$10$^{4}$ stars (dG92), similar to the better-studied 
Orion complex.  As in Orion, several rich associations have formed in Sco-Cen over a 
$\sim$15 Myr period, producing the US-UCL-LCC subgroups, as well as many
smaller, peripheral molecular clouds and dispersed clusters.  
These small, unbound stellar groups formed in either the
progenitor Sco-Cen GMC, or in short-lived molecular clouds formed by the superbubbles. 
These groups are partly or fully 
evaporated today and are now seen as dispersed collections of pre-MS stars 
that we label as the TW Hya Association, $\eta$ Cha cluster, $\beta$ Pic group, etc. 
Such stellar dispersal over 10s-100\,pc over $\sim$10 Myr
is expected if the stars inherited their space motions from clouds which were 
endowed momentum either from (1) supersonic turbulence in GMCs or (2) 
formation in molecular clouds associated with expanding superbubbles. 

We believe that this scenario provides a convincing explanation 
for the presense of many (or most) of the isolated, extremely 
young ($\sim$10 Myr-old) stars in the solar neighborhood.  It is 
supported by the approximate agreement between the kinematic and 
nuclear ages.
Observationally, we see the ``process'' continue today in the
``150\,pc'' molecular clouds in the vicinity of the Sco-Cen OB subgroup. 
The key to testing these ideas is higher quality 
astrometric data: more precise parallaxes, proper motions, and radial 
velocities for all nearby pre-MS stars.  Except for radial velocities 
which can be obtained from ground-based telescopes, missions like {\it FAME}, 
{\it DIVA} and {\it GAIA} are critical to this effort.

\acknowledgements

Many thanks to W. A. Lawson for long discussions of these issues. E. E. M. 
would like to thank the SIRTF Legacy Science program (JPL Contract 1224768), 
and NASA Ames for support.
E. D. F.'s research is supported in part by NASA contract 
NAS 8-38252 and NAG 5-8422. Thanks go to J. Liebert, M. Meyer, and
J. Moustakas for
critiquing the draft. This research made use of the ADS (NASA), 
and the SIMBAD and Vizier services (CDS).

\end{document}